\documentclass[12pt]{article}

\usepackage{amsmath}
\usepackage{amssymb}

\def\rdots{\mathinner{\mkern1mu\raise1pt\vbox{\kern1pt\hbox{.}}\mkern2mu
   \raise4pt\hbox{.}\mkern2mu\raise7pt\hbox{.}\mkern1mu}}
\newcommand{\Z}{{\rm Z\kern-.35em Z}}
\newcommand{\bP}{{\rm I\kern-.15em P}}
\newcommand{\Q}{\kern.3em\rule{.07em}{.65em}\kern-.3em{\rm Q}}
\newcommand{\R}{{\rm I\kern-.15em R}}
\newcommand{\h}{{\rm I\kern-.15em H}}
\newcommand{\C}{\kern.3em\rule{.07em}{.65em}\kern-.3em{\rm C}}
\newcommand{\T}{{\rm T\kern-.35em T}}

\newcommand{\be}{\begin{equation}}
\newcommand{\ee}{\end{equation}}

\newcommand{\la}{\lambda}
\newcommand{\Lam}{\Lambda}

\newcommand{\pa}{\partial}

\newcommand{\ra}{\rightarrow}

\newcommand{\al}{\alpha}

\begin{document}

\openup 1.5\jot
\centerline{Computation of Terms in the Asymptotic Expansion of Dimer $\la_d$ }

\centerline{for High Dimension}

\bigskip

\bigskip

\vspace{1in}
\centerline{Paul Federbush}
\centerline{Department of Mathematics}
\centerline{University of Michigan}
\centerline{Ann Arbor, MI 48109-1043}
\centerline{(pfed@umich.edu)}

\vspace{1in}

\centerline{\underline{Abstract}}

\ \ \ \ \ The dimer problem arose in a thermodynamic study of diatomic molecules, and was abstracted into one of the most basic and natural problems in both statistical mechanics and combinatoric mathematics.  Given a rectangular lattice of volume $V$ in $d$ dimensions, the dimer problem loosely speaking is to count the number of different ways dimers (dominoes) may be laid down in the lattice (without overlapping)  to completely cover it.  Each dimer covers two neighboring vertices.  It is known that the number of such coverings is roughly $e^{\la_d V}$ for some constant $\la_d$ as $V$ goes to infinity.  Herein we present a mathematical argument for an asymptotic expansion for $\la_d$ in inverse powers of $d$, and the results of computer computations for the first few terms in the series.  As a glaring challenge, we conjecture no one will compute the next term in the series, due to the requisite computer time and storage demands.

\vfill\eject

For simplicity and to be specific we consider a periodic cubical lattice $\Lam$ with the edge length even.  Then the number of different coverings by dimers has as dominant behavior, in the $V \ra \infty$ limit,
\be	e^{\la_d V} \ee
for a d-dimensional lattice, where the number of vertices we denote by both $N$ and $V$.  The asymptotic relation 
\be	\la_d \sim \frac 1 2 \; ln(2d) - \frac 1 2 \ ,   \ee
was exactly presaged over 40 years ago by the results of [1], and this work was largely the basis for a complete proof of (2) a decade later in [2].  While in two dimensions $\la_2$ is known exactly, [3], [4], in dimensions higher than two it is computationally difficult to get good bounds for $\la_d$.  In a series of reports [5], [6], [7] we have found a mathematical argument for a full asymptotic expansion for $\la_d$
\be	\la_d \sim \ \frac 1 2 \ ln(2d) - \frac 1 2 + \frac{c_1}d + \frac{c_2}{d^2} + \cdots \   \ee
and in [8] we presented the result of computer computations of the first three $c_i$ leading to
\be	\la_d \sim \frac 1 2 \; ln(2d) - \frac 1 2 + \frac 1 {8} \frac 1 d + \frac5{96} \; \frac1{d^2}  + \frac 5{64}\;  \frac1{d^3} + \cdots. \ee
In the Appendix we show the consistency of (4) with known rigorous bounds on $\la_d$.  We there also make some ``intelligent guesses" for $\la_2$ and $\la_3$ based on the present theory, whose values further increase our confidence.  We proceed to clarify and expand upon some of the developments in [5] - [8].

In $d$ dimensions there are $d$ `kinds' of dimers, each kind oriented in one of the $d$ lattice directions.  Each kind of dimer may be `located' at some place on the lattice.  We generalize this situation as follows.  A `located tile' is a two element subset of the lattice.  A `tile' is an equivalence class of located tiles, with equivalence given by setting two subsets equivalent if one is a translation of the other.  (A `tile' is a generalization of a `kind of dimer'.)  Notice that tiles need not be connected.

We consider tiles with a `weighting', a function on tiles.  We normalize the weightings we consider, by requiring, if $g$ is the weighting function,
\be	\sum_t \ g(t) = 1/2 \; ,	\ee
the sum over all tiles, $t$.  We let $f$ be the weighting function given by
\be
f(t) = \Bigg\{ 
\begin{array}{cc} 
\frac 1{2d} & {\rm if} \ \ t \ \ {\rm is \ a \ dimer} \\
 \\
0 & \ {\rm otherwise}
\end{array}
\ee
that clearly satisfies the normalization condition, (5).

A `tiling' $T_i$ of $\Lam$ is a set of two element subsets of $\Lam$,
\be	T_i = \left\{ s^i_1, s^i_2, ..., s^i_{N/2} \right \}  \ee
where the $s^i_k$ are disjoint.

We now realize the sum over all possible dimer covers of $\Lam$, the goal of our study, as
\be	(2d)^{N/2} \ Z	\ee
with
\be	Z = \sum_{T_i} \prod_{s_\al \in T_i} f(\bar s_\al)    \ee
where the sum is over all tilings of $\Lam$; the product over the $f's$ selects those tilings in which all the tiles employed are dimers.  The bar over a subset indicates the equivalence class of the subset as defined previously.

We let $f_0$ be a constant function on tiles with value $\frac 1{N-1}$, to satisfy the normalization condition (5).  We write
\be	f = f_0 + (f - f_0) \equiv f_0 + v,  \ee
$f, f_0$, and $v$ all functions on tiles.  $Z$ becomes
\begin{eqnarray}
	Z &=& \sum_{T_i} \prod_{s_\al \in T_i} \Big( f_0 + v(\bar s_\al) \Big) \\
	Z&=& Z_0 + Z_1 + Z_2 + . . . 	
\end{eqnarray}
having expanded $Z$ in powers of $v$.

We note
\be		e^{N\la_d} = (2d)^{N/2} \; Z	\ee
Here $\la_d$ is understood a function of $N$, the usual $\la_d$ the infinity limit.  Or equivalently
\be		\la_ d = \frac 1 2 \; ln(2d) + \frac 1 N \ ln \; Z	\ee
If one replaces $Z$ by $Z_0$, the mean-field approximation in a natural nomenclature, one gets taking the infinity limit
\be   \la_d \cong \frac 1 2 \ ln(2d) - \frac 1 2  = \frac 1 2 \; ln(2d) + \lim_{N\ra \infty} \frac 1 N \; \ln \; Z_0  \ee
by an easy calculation.  Thus equation (2) arises from the mean-field approximation, keeping only $Z_0$ in (12).

We return to (12) and introduce some convenient notations.
\begin{eqnarray}
Z &=& Z_0 Z^*  \\
Z^* &=& 1 + Z^*_1 + Z^*_2 + ..... \\
Z^*_i &=& Z_i \big/ Z_0
\end{eqnarray}
There is a natural factorization of $Z_i$ into a contribution from the factors of $v$ in (11) which we call $\bar Z^*_i$ and the factors of $f_0$ in (11) which we call $\beta(N,i) Z_0$ so that
\be   Z^*_i = \beta(N,i) \bar Z^*_i.	\ee
A tedious calculation shows
\be   \beta(N, jN) \sim e^{N\big[ \big(\frac{1-2j}{2}\big) ln(1-2j)+j\big]}  \ee
for large $N$, a result we later need.

We let $\tilde Z^*$ be $Z^*$ with $\beta(N,i)$ replaced by 1,
\be	\tilde Z^* = 1 + \bar Z^*_1 + \bar Z^*_2 + . . . 	\ee
where a detailed specification of $\bar Z^*_i$ is given by
\be
\bar Z^*_i = \frac 1{i!} \begin{array}[t]{c}
{\displaystyle\sum} \\
{\scriptstyle{s_1,s_2,...,s_i }} \\
{\scriptstyle {{\rm disjoint }}}
\end{array}
\prod^i_{\al = 1} v(\bar s_\al)
\ee
Now referring to [9] we may write a cluster expansion for $\tilde Z^*$ 
\be	ln \ \tilde Z^* = \sum_s \frac 1{s!} \ J_s	\ee
\be	J_s = \sum_{s_1,s_2,...,s_s} \ v(\bar s_1) ... v(\bar s_s) \psi'_c(s_1,s_2,...,s_s)	\ee
where we may identify equation (2.5a) of [9] with $\tilde Z^*$, and (23,), (24) with equation (2.7) of [9].  The located tiles in the sum of (24) are forced to overlap so they cannot be divided into two disjoint sets.  $\psi '_c$ is a numerical factor depending on the overlap pattern.  To make our computations mathematically rigorous it will be necessary to study the convergence properties of sums such as in (23).  At present this appears very difficult, and we by no means see yet a clear route to a proof.  It is a challenging problem for the mathematical physicist.  

It is easy to show $J_1 = 0$ and it is proven in [6] that
\be		J_s = \frac{C_{s,r}}{d^r} + \frac{C_{s,r+1}}{d^{r+1}} + \cdots + \frac{C_{s,s-1}} {d^{s-1}} \  .  	\ee
with $r \ge s/2$.  We also find it convenient to define
\be  N \bar J_i = (1/i!)J_i	\ee

From (17), (19), (21), (23) one gets
\be  Z^* = \sum_{\al_1,...,\al_{s+1}} \beta\big(N, \Sigma\; i\; \al_i \big) \bar J^{\al_1}_1 \cdots \bar J^{\al_{s+1}}_{s+1} \ 
\frac{N^{\Sigma \; \al_i }}{\al_1! \cdots \al_{s+1} !}  \ee
We approximate the sum in (27) by its largest term, in the limit $N \ra \infty$.  If all the $\bar J$'s are positive this is a reasonable way to extract the dominant asymptotic limit.  In [7] an argument is given that our results will hold even if some of the $\bar J$'s are negative.   

We differentiate with respect to the $\al$'s to find the largest term in (27), using (20) to treat the $\beta$ factor and find the following equations, upon scaling $\al_i \ra \frac 1 N \; \al_i$.

\be	ln \; \al_k = ln \; \bar J_k + \frac \pa{\pa \al_k} \left[ \left( \frac{1-2 \Sigma i \al_i}{2} \right) ln (1 - 2\Sigma \; i \al_i) + 
\Sigma \; i \al_i  \right]  \ee
or
\be	\al_k = \bar J_k e^{F_k(\al's)}   \ee
where (28) and (29) define the $F_k$.  Extracting the term in the sum of (27) with the $\al_k$ of (27) equal $N$ times the $\al_k$ of (29) we get 
\be
Z^* \ \sim \ e^{N \Big\{ - \Sigma \;\al_i\; F_i \;+\; \Sigma \; \bar J_i \; e^{F_i}\;+\; \frac{1-2\Sigma \;i \;\al_i}{2} \; ln \big(1-2\Sigma \; i \;\al_i\big) + \Sigma i \al_i  \Big\} } .
\ee
Using (28) and (29) we can expand the exponent in (30) into a power series in the $\bar J$'s.  From (14), (15), and (16) then we can find $\la_d$ as a formal power series in the $\bar J$'s.  Once the $\bar J$'s are known, one can extract a formal power series in the inverse powers of $d$ (see (25)), and our formula (4).  Everything rests on finding the $\bar J$'s\; !

All computer computations were done using maple in integer arithmetic.  We note $\bar J_i $ is a function of  dimension.  We computed $\bar J_1, \bar J_2, ... , \bar J_6$ in dimensions 1, 2, and 3 by computing the  expression in (24).  By a simple device one may replace the $v$'s in (24) by $f$'s, and afterwards recover the  original expression.  Thus the most complicated computation of these involved placing down six dimers in three dimensions, in all possible ways where overlaps make it impossible to disconnect into two disjoint subsets of  dimers.  From the form of $\bar J_i$ as given in (25) we note this determines $\bar J_i,  i = 1,...,6$ in all dimensions.  To compute the $c_4$ term in (3) one would require $\bar J_7$ and $\bar J_8$ in arbitrary dimensions.  This would entail computations in dimensions 1, 2, 3, and 4, the most complex setting down eight dimers in four dimensions.  As another measure of difficulty of computing $c_4$, we note it took approximately 10 seconds of computer time to compute $c_2$, and two weeks to compute $c_3$ .... and we are using an exponential time algorithm.  Good luck to the hearty soul attempting this computation. 
 
The computer results were first presented in [8].  The following table includes the computations of the $\bar J$'s 
\[
\begin{tabular}{c|c|c|c|}
   & $d=1$  &  $d=2$ & $d=3$  \\ \hline
 & & &  \\
$\bar J_1$ & 0 & 0 & 0  \\
  & & & \\
$\bar J_2$ & 1/8 & 1/16 & 1/24  \\
  & & & \\ 
$\bar J_3$ & 1/12 & 1/48 & 1/108  \\
 & & &  \\ 
$\bar J_4$ & - 3/64 & - 9/512 & - 5/576 \\
 & & & \\ 
$\bar J_5$ & - 13/80 & - 23/1280 & - 11/2160 \\
  & & & \\ 
$\bar J_6$ & -19/192 & 25/3072 &  175/46656
\end{tabular}   \]
From these results there follows from (25) the following expressions for $\bar J_i$
\begin{eqnarray}
\bar J_1 &=& 0 \\
\bar J_2 &=&  \frac 18 \; \frac 1 d \\
\bar J_3 &=& \frac 1{12}\; \frac1{d^2} \\
\bar J_4 &=&-\; \frac 3{32} \; \frac1{d^2} + \frac3{64} \; \frac 1{d^3} \\
\bar J_5 &=&  - \; \frac 1 8 \; \frac 1{d^3} - \frac 3{80} \; \frac 1 {d^4} \\
\bar J_6 &=&  \frac 7{48}  \frac1{d^3} -\frac 5{64} \frac 1{d^4} -  \frac 1 6 \frac1{d^5}
\end{eqnarray}

\bigskip
\bigskip

\centerline{ Appendix}

We first study what rigorous upper and lower bounds on $\la_d$, for large $d$, have to say about our
expansion (4).  For an upper bound we have from eq. (9) and eq. (10) of [2] that
\be	\la_d < \frac 1 2 \ ln(2d) - \frac 1 2 + \frac 1{8d} \ ln(4\pi d) + \frac 1{96d^2}.  \ee
Clearly an asymptotic expansion such as (3) with any value for the $c$'s is consistent with (37), the $\frac 1 d \; ln(d)$ dominating inverse powers of $d$ asymptotically.

For a lower bound estimate one has from [10], and [11] Formula (5.4), [12] Equation (6.8)
\be \la_d \ge \frac 1 2 \Big( (2d-1) ln (2d-1) - (2d-2) ln(2d) \Big)	\ee
from which follows
\be	\la_d \ge \frac 1 2 \ ln(2d) - \frac 1 2 + \sum^\infty_{i=1} \ \frac 1{(2i)(i+1)(2d)^i} \ee
\be	\ge \frac 1 2 \ ln(2d) - \frac 1 2 + \frac 1 8 \; \frac 1 d + \frac 1{48} \; \frac 1{d^2}.	\ee
For (3) to be consistent with (40) one must have
\be	c_1 \ge \frac 1 8 \ee
and if $c_1 = 1/8$ (as it does) then
\be	c_2 \ge \frac 1{48}.	\ee
With $c_1 = 1/8$ and $c_2 = 5/96$ both inequalities, (41) and (42), are satisfied.  A minimal test for the theory passed.  The estimate (39) was pointed out to us in a personal communication by Shmuel Friedland.

In extracting an asymptotic expansion in inverse powers of $d$ from (28), (29), (30) we needed two $\bar J_i$ values for each inverse power of $d$. (e.g. we needed $\bar J_1, ..., \bar J_6$ to derive $c_1, ..., c_3$.)  In [8] and [13] we introduced a formal parameter $x$ (ultimately set equal to 1), replaced $\bar J_i$ by $\bar J_i \; x^{i-1}$ in (28) - (30) and found an expansion in powers of $x$, this seemed a possible improvement.  (The expansion in powers of $x$ is formally somewhat like an expansion in powers of $1/\sqrt{d}$.)

We then viewed the expansion in powers of $x$, as an asymptotic expansion.  Follo
wing the notation of [8] and [13] we let $B_k$ be the partial sum for $\la_d$ keeping powers of $x$ 
through $x^k$, then setting $x$ to be one.  We let $B_g$ and $B_{g+1}$ be the two successive $B_k$ with minimum value for $|B_{g+1} - B_g|$.  We set
\be	a = \frac 1 2(B_g + B_{g+1})	\ee
and 
\be	b = |B_g - B_{g+1} | \ee
and took our estimate for $\la_d$ to be
\be	\la_d = a \pm b   \ee
by the rule of thumb I learned as an undergraduate physics major.  For $d=2$ and $d=3$ this yielded
\be	\la_2 = .296 \pm .007	\ee
and
\be	\la_3 = .453 \pm .001.	\ee
$\la_2$ is known exactly as
\be	\la_2 = .2915 ....  \ee
by [3] and [4], and
\be	.440075 \le \la_3 \le .457547	\ee
by [15], an improvement on [14].

Viewing the consistency of (46) with (48), of (47) with (49), and the validity of (41), and (42); we may associate a casual probability of $\frac 1{30}, \frac 1{50}, \frac 1 2, \frac 1 2$ for each of these four consistencies to occur at random.  {\it So we might say the odds are 6000 to one our theory is correct.}  While we may have presented this argument in a frivolous manner, some encouraging points are being made.  (Using the consistency of (46) - (49) as an argument for the theory was suggested to me by Mihai Ciucu.)

\underline{Acknowledgment}.  I would like to thank Shmuel Friedland, Gordon Slade and Nathan Clisby for support along the way.

\bigskip

\vfill\eject

\centerline{\underline{ References}}

\bigskip

\begin{itemize}
\item[[1]]  J. M. Hammersley, ``An Improved Lower Bound for the Multidimensional Dimer Problem", Proc. Cambridge Philos. Soc. {\bf 64}, (1968), 455-463. 
\item[[2]] Henryk Minc, ``An Asymptotic Solution of the Multidimensional Dimer Problem, Linear and Multilinear Algebra", 1980, {\bf 8}, 235-239.
\item[[3]] E.M. Fisher, ``Statistical Mechanics of Dimers on a Plane Lattice", Phys. Rev. 124 (1961), 1664-1672.
\item[[4]] P.W. Kasteleyn, ``The Statistics of Dimers on a Lattice", Physica 27 (1961), 1209-1225.
\item[[5]] Paul Federbush, ``Hidden Structure in Tilings, Conjectured Asymptotic Expansion for $\la_d$ in Multidimensional Dimer Problem", arXiv:0711.1092V9[math-ph].
\item[[6]] Paul Federbush, ``Dimer $\la_d$ Expansion, Dimension Dependence of $\bar J_n$ Kernels", arXiv:0806.1941V1[math-ph].
\item[[7]] Paul Federbush, ``Dimer $\la_d$ Expansion, A Contour Integral Stationary Point Argument",
arXiv:0806.4158V1[math-ph].
\item[[8]] Paul Federbush, ``Dimer $\la_d$ Expansion Computer Computations", arXiv:0804.4220V1 [math-ph].
\item[[9]] David C. Brydges, ``A Short Course in Cluster Expansions" in ``Phenomenes Critiques, Systems Aleatoires, Theories de Gauge, Part I, II" (Les Houches, 1984), 129-183, North Holland, Amsterdam, 1986.
\item[[10]] A. Schrijver, ``Counting 1-factors in Regular Bipartite Graphs", J. Comb. Theory {\bf B72} (1998), 122-135.
\item[[11]] S. Friedland and U.N. Pelod, ``Theory of Computation of Multidimensional Entropy with an Application to the Monomer-Dimer Problem", Advances of Applied Math. {\bf 34} (2005), 486-522.
\item[[12]] S. Friedland, ``Multidimensional Capacity, Pressure and Hausdorff Dimension", 
Mathematical Systems in Biology, Communications, Computations, and Finance,  Editors: J. Rosenthal and D. Gillian, IMA vol. 134 (2003), 183-222.
\item[[13]] P. Federbush, ``Dimer $\la_3 = .453 \pm .001$ and Some Other Very Intelligent 
Guesses", arXiv:0805.1195V1 [math-ph].
\item[[14]] Mihai Ciucu, ``An Improved Upper Bound for the 3-dimensional Dimer Problem", Duke. Math. J. {\bf 94}
(1998), 1-11.
\item[[15]] S. Friedland, E. Krop, P.H. Lundow, and K. Markstrom, ``Validations of the Asymptotic Matching Conjectures", Journal of Statistical Physics, {\bf 133} (2008) 513-533.
\end{itemize}

\end{document}